\documentclass[lettersize,journal]{IEEEtran}
\usepackage{amsmath,amsfonts,amssymb}
\usepackage{algorithm}
\usepackage{algorithmic}
\usepackage{array}
\usepackage[caption=false,font=normalsize,labelfont=sf,textfont=sf]{subfig}
\usepackage{textcomp}
\usepackage{stfloats}
\usepackage{url}
\usepackage{verbatim}
\usepackage{graphicx}
\def\BibTeX{{\rm B\kern-.05em{\sc i\kern-.025em b}\kern-.08em
		T\kern-.1667em\lower.7ex\hbox{E}\kern-.125emX}}
\usepackage{balance}

\usepackage{amsthm} 

\newtheorem{assumption}{Assumption}
\newtheorem{remark}{Remark}
\newtheorem{problem}{Problem}
\newtheorem{definition}{Definition}
\newtheorem{theorem}{Theorem}

\usepackage{color}
\usepackage[hidelinks]{hyperref}
\usepackage{cite}
\usepackage{algorithm}
\usepackage{bm}

\begin{document}
	\title{Contouring Error Bounded Control for Biaxial Switched Linear Systems}
	\author{Meng Yuan, \emph{Member,~IEEE}, Ye Wang, \emph{Member,~IEEE}, Chris Manzie, \emph{Senior Member,~IEEE}, Zhezhuang Xu, \emph{Member,~IEEE}, Tianyou Chai, \emph{Life Fellow,~IEEE}
		
		\thanks{This work is supported in part by European Union (EU)-funded Marie Sklodowska-Curie Actions (MSCA) Postdoctoral Fellowship under grant number 101110832, in part by the Australian Research Council via the 2022 Discovery Early Career Researcher Award (DECRA) under grant number DE220100609, in part by National Natural Science Foundation of China under projects number 61973085 and 61991404, and in part by the Research Program of the Liaoning Liaohe Laboratory under number LLL23ZZ-05-01. 
			
		Meng Yuan is with the Department of Electrical Engineering at Chalmers University of Technology, Gothenburg 41296, Sweden. ({\tt\small meng.yuan@ieee.org})
			
		Ye Wang and Chris Manzie are with the Department of Electrical and Electronic Engineering, The University of Melbourne, Parkville, VIC 3010, Australia. 

        Zhezhuang Xu is with the College of Electrical Engineering and Automation, Fuzhou University, Fuzhou 350108, China. 
        
		Tianyou Chai is with the State Key Laboratory of Synthetical Automation for Process Industries, Northeastern University, Shenyang 110819, China. 
	}
}

	\maketitle

	\begin{abstract}

    Biaxial motion control systems are used extensively in manufacturing and printing industries. To improve throughput and reduce machine cost, lightweight materials are being proposed in structural components but may result in higher flexibility in the machine links. This flexibility is often position dependent and compromises precision of the end effector of the machine. To address the need for improved contouring accuracy in industrial machines with position-dependent structural flexibility, this paper introduces a novel contouring error-bounded control algorithm for biaxial switched linear systems. The proposed algorithm utilizes model predictive control to guarantee the satisfaction of state, input, and contouring error constraints for any admissible mode switching. In this paper, the switching signal remains unknown to the controller, although information about the minimum time the system is expected to stay in a specific mode is considered to be available. The proposed algorithm has the property of recursive feasibility and ensures the stability of the closed-loop system. The effectiveness of the proposed method is demonstrated by applying it to a high-fidelity simulation of a dual-drive industrial laser machine. The results show that the contouring error is successfully bounded within the given tolerance.

	\end{abstract}
	

\section{Introduction}

\IEEEPARstart{C}{ontrol} of biaxial machining systems, such as laser cutters and water-jet machines, involves precise and coordinated movement of the end-effector in two-dimensional space \cite{Yuan2020}. The contouring error, defined as the minimum distance between the end-effector position and desired path, directly quantifies the machining accuracy of manufactured products \cite{tang2013multiaxis,yao2011orthogonal,qin2017contour}.
	
While minimizing contouring error remains a fundamental objective in biaxial motion control, ensuring bounded contouring error offers a comprehensive solution that maintains quality assurance, operational safety and process reliability~\cite{chen2019contour}. It guarantees that the manufacturing process not only achieves high precision but also operates within well-defined and acceptable error limits.
	
Achieving bounded contouring error poses practical challenges, particularly in contemporary applications where manipulators are designed for faster movements using lighter materials. While reduced weight enhances agility, it introduces position-dependent structural flexibility, making the movement of both axes coupling. This results in a system with high nonlinearity, which makes the controller design problem challenging \cite{symens2004gain}.

Modelling the dynamics with sufficient fidelity, and incorporating the model into the controller has been shown previously to improve contouring performance, ranging from cantilever beams to dual-drive machines \cite{teo2007dynamic, lu2012two, li2018modeling}. In \cite{teo2007dynamic}, the Lagrangian equations were used to describe the dynamics of the gantry machine system. In \cite{lu2012two}, an Euler-Bernoulli beam model was presented with the truncated mode method to represent the dynamics of a machine with the cantilever beam structure. To capture the coupling in both axes, virtual springs are introduced in modeling process to approximate the effect of ball bearings between motors and linear guides \cite{li2018modeling}.

The presence of structural flexibility introduces discrepancies between the end-effector and actuator positions, leading to adverse effects on the accuracy of manufactured products, especially in high-acceleration scenarios. To address this challenge, detailed modeling becomes essential in controller design. One effective method to enhance modeling accuracy is by incorporating the rotation angle as an additional state, resulting in interconnected and complex control dynamics~\cite{chen2019integrated}. To simplify the subsequent controller design process, the nonlinear system model is normally linearized at different operating points and accommodating bounded disturbances. This simplification process leads to a switched linear system with a state-dependent switching signal. While the time-dependent reference is typically provided in advance to offer information on the minimum duration the system can stay in each mode, the exact switching signal remains unknown to the controller. 
   
To minimize the contouring errors when a reference trajectory is specified, the cross-coupled control (CCC) was first developed to determine control action by sending the weighted contour error equally to axis controllers and forcing the axial tool position onto the path \cite{koren1980cross}. However, the constant compensator value and cross-coupling gains make the conventional CCC ineffective in dealing with nonlinear contours and can also lead to oscillation for linear contours when steady-state error tends to zero \cite{koren1991variable}. To deal with this drawback in the conventional CCC, methods including variable gain approach \cite{koren1991variable} and sliding mode control combined with CCC \cite{kuang2020precise} were introduced. In \cite{huo2012improving}, several CCC methods were reviewed and each method combines a specific contour error estimation algorithms and a set of control laws for directly reducing the contouring error.
	
The coordinate transformation-based control is an alternative method to improve the contouring performance, where decoupled controllers are designed to reduce the transformed errors \cite{yao2011orthogonal}. In \cite{chiu2001contouring}, the error dynamics are represented in tangential and normal components, and the controller is designed to reduce the normal component of errors. Instead of estimating the contouring error, an equivalent error-based contouring control was presented in \cite{chen2007contouring}. The reduction of the equivalent error leads to a reduced contouring error. While coordinate transformation-based control schemes simplify the controller design, the contouring performance deteriorates significantly when dealing system with nonlinearities \cite{yang2019novel}. In addition, the state, input and contouring error-related constraints are not considered in the above-discussed methods. 
	
Model predictive control (MPC), due to its advanced capabilities in real-time trajectory tracking, has become a prevalent choice in industrial motion control \cite{van2011model,di2018cascaded,li2022finite,picotti2022nonlinear}. Its inherent advantages in handling complex systems by explicitly addressing state, input, and performance constraints make it particularly well-suited for the intricate requirements of biaxial contouring control. In \cite{lam2012model}, a time-varying MPC was proposed, aiming to concurrently maximize contouring accuracy and minimize traverse time. The design of this controller also takes into account factors such as friction and other disturbances by incorporating a disturbance model in the prediction model of MPC. In \cite{zhang2018generalized}, a generalized MPC was developed based on a unified model to simplify the modeling process. This approach introduces tracking error, contouring error, and input increment into the cost function, enabling the determination of optimal control through multi-step prediction.

For linear time-invariant systems, the constraints of states and contouring errors can be guaranteed by enforcing the states within a designed control invariant (CI) set \cite{yuan2019modelling}. However, satisfying constraints for any possible switching sequence is challenging considering the feasibility of MPC can be lost when dynamics change. To ensure the constraints for the switched linear system, an MPC-based algorithm for a discrete-time switched linear system with known switching signals was proposed \cite{bridgeman2016stability}. The terminal cost and constraints that depend on the terminal mode of the system are designed to ensure persistent feasibility and asymptotic stability under dwell-time restrictions. Later, the algorithm was extended to an unknown mode switching case in \cite{danielson2019necessary}, where the switching signal is subject to dwell time and mode transition restrictions.

In this work, we aim to propose a contouring error-bounded control algorithm tailored for biaxial industrial systems with position-dependent flexibility. The main objective is to ensure that the end-effector can follow a desired trajectory while adhering to a prescribed contouring error tolerance. If the reference is not admissible, the controller can steer the system to the closest admissible points. To achieve this, we initiate the discussion by considering an approximated switched linear system. hen, we propose an algorithm for computing switch CI sets, incorporating set propagation and mode transition restrictions that can guarantee state, input and contouring error constraints during the manufacturing process. To tackle the challenge of computing switch CI sets from non-compact feasible sets, we develop an optimization-based algorithm for approximating feasible sets while reducing the computational complexity from an implementation viewpoint. Theoretical results on the recursive feasibility and closed-loop stability of the proposed method are presented. Finally, the efficacy of the proposed approach is validated through comprehensive testing on a high-fidelity model of an industrial laser machine.

\emph{Notation}: $\mathbb{R}$, $\mathbb{Z}{+}$, and $\mathbb{N}$ denote the sets of real, positive integer and natural numbers, respectively. The closed integer set with a range from $m$ to $n$ is denoted by $\mathbb{Z}_{[m,n]}$. For a matrix $\mathbf{X} \in \mathbb{R}^{n\times n}$, $\mathbf{X} \succ 0$ and $\mathbf{X} \succeq 0$ signify positive and semi-positive definiteness, respectively. A diagonal matrix with main diagonal elements $a$ and $b$ is represented as $\mathbf{X} = \mathrm{diag}(a,b)$. The vector $\mathbf{x}(k)$ denotes the value of $\mathbf{x}$ at the sampling instant $k$, and the vector $\mathbf{x}(i|k)$ stands for the predicted value of $\mathbf{x}$ at time instant $k+i$ based on the data sampled at time $k$. For a vector $\mathbf{x} \in \mathbb{R}^{n}$ with a matrix $P \succ 0 \in \mathbb{R}^{n \times n}$, we use $\| \mathbf{x} \|_2$ and $\| \mathbf{x}\|_{P}$ to denote 2-norm and weighted 2-norm, respectively. 
	
\section{Problem statement} \label{sec:prob_state}

In this section, we establish the necessary preliminaries for the problem formulation, covering aspects from the definition of a switched linear system to contouring error. 
	
\subsection{Switched linear system}

\begin{figure}[tbp]
	\centering
	\subfloat[]
	{	\centering
		\label{fig:AM_LM}
		\includegraphics[width = 0.9\hsize]{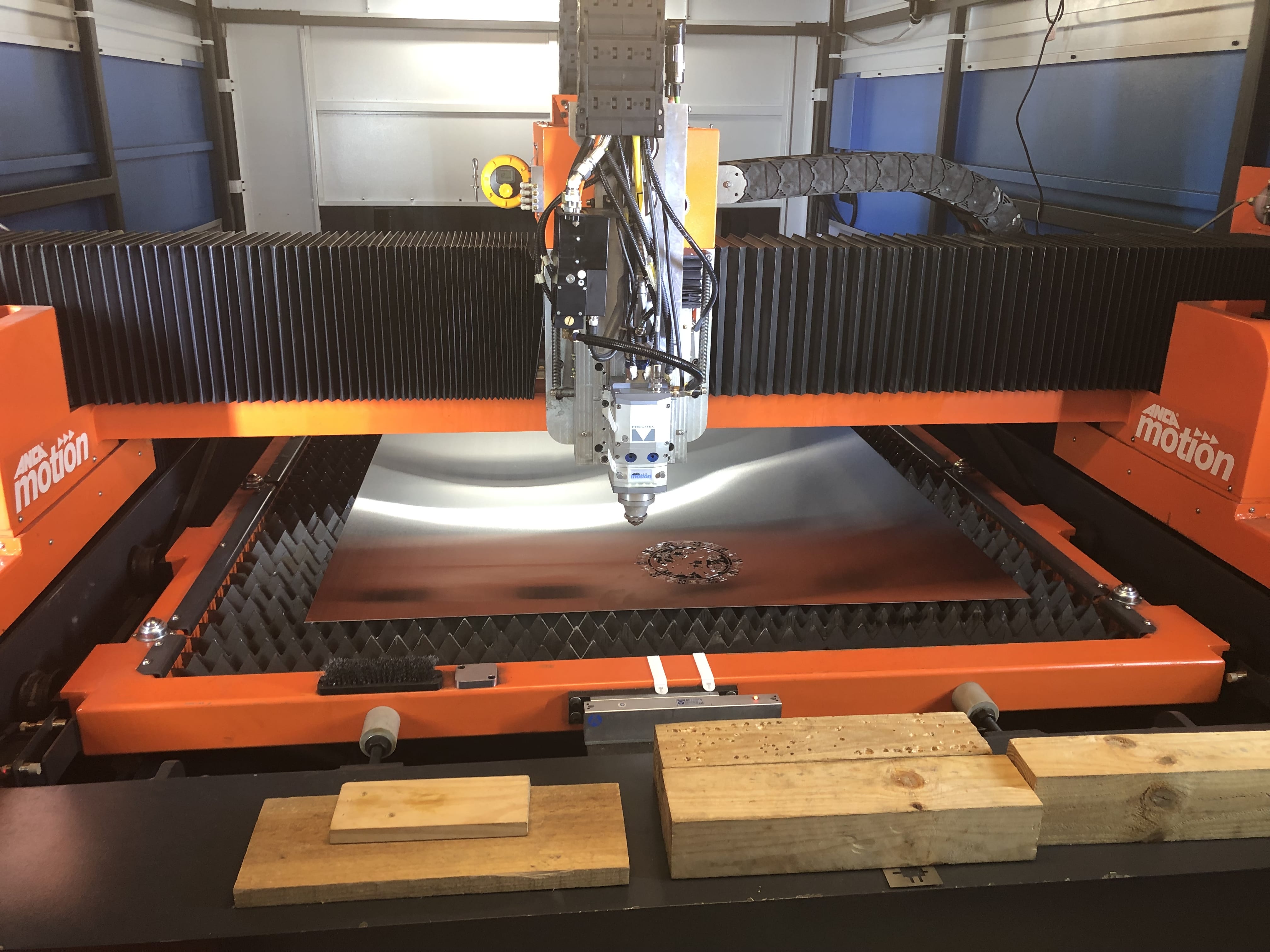}
	}
	\\
	\subfloat[]
	{	\centering
		\label{fig:laser_schematic}
		\includegraphics[width = 0.9\hsize]{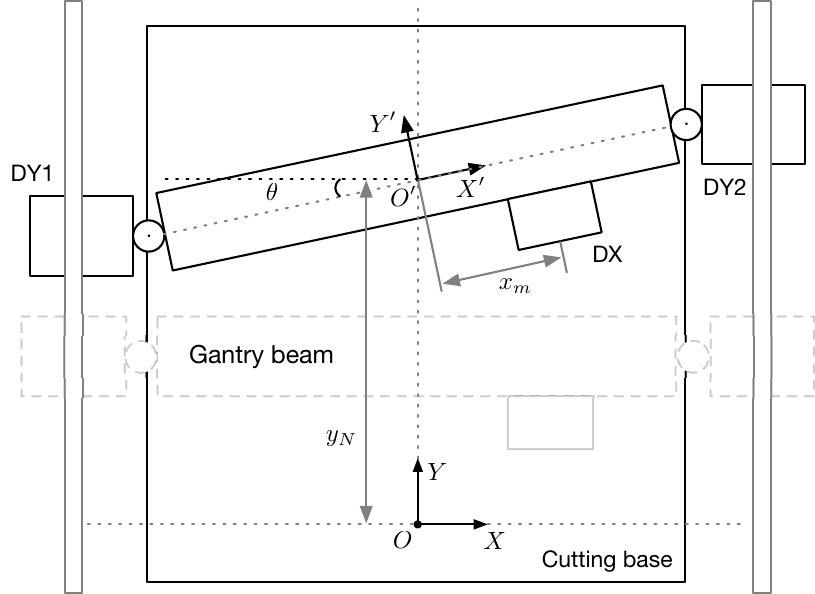}
	}
	\caption{Industrial dual drive machine: (a) photo of industrial laser machine from our partner ANCA Motion; (b) schematic diagram of dual drive machine (top view).\label{fig:laser}}
\end{figure}

The biaxial industrial system generally consists of components such as end-effector motors and dual drive motors as shown in Fig.~\ref{fig:AM_LM}. The inherent structural flexibility in biaxial systems introduces the coupling of lateral and longitudinal motions, with the rotation angle serving as an additional state when describing the system dynamics. The behavior of system under steady-state conditions (dotted grey line) and moving motion (solid black line) are illustrated in Fig.~\ref{fig:laser_schematic}, revealing synchronization imperfections of dual drives.

As the end-effector moves, the system dynamics become state-dependent, forming a direct correlation between the system behavior and the end-effector position. Through linearization at various operating points, an approximation of initial nonlinear system is represented by a switched discrete-time linear system as:
\begin{subequations}\label{eq:switched_system}
	\begin{align}
			\mathbf{x}(k+1) &= A_{\sigma(k)}\mathbf{x}(k) + B_{\sigma(k)}\mathbf{u}(k), \\
			\mathbf{y}(k) &= C_{\sigma(k)}\mathbf{x}(k),
	\end{align}
\end{subequations}
where $\mathbf{x} \in \mathbb{R}^{n_{x}}$ represents the state, $\mathbf{u} \in \mathbb{R}^{n_{u}}$ is the input. The output is the end-effector position in the biaxial system, i.e. $\mathbf{y} = \left[ x_{e}, y_{e} \right]^{\top} \in \mathbb{R}^{2}$. The switching signal $\sigma: \mathbb{Z} \rightarrow \mathcal{M}$ is a piece-wise constant function with time $k$ that switches the system dynamics between a finite number of modes, $m \in \mathcal{M}  = \{1,\cdots, M\}$.
	
For the biaxial contouring problem, the switching signal is state-dependent, and the system mode $\sigma(k)$ is known at time instant $k$. The state and input are required to stay within polytopic sets:
\begin{equation}
	\mathbf{x} \in \mathcal{X}, \mathbf{u} \in \mathcal{U}. \label{eq:cons_sys}
\end{equation}

\begin{assumption}\label{ass:stab_obs}
	The pair $(A_{m}, B_{m})$ is stabilizable and the pair $(A_{m}, C_{m})$ is observable for all $m\in \mathcal{M}$. 
\end{assumption}

  The motion of the end-effector is expected to track a desired curve denoted by a finite number of reference sequence $\bar{\mathbf{r}} = \{ \mathbf{r}_{1},\cdots,\mathbf{r}_{j},\cdots,\mathbf{r}_{N_{r}}\}$ with $\mathbf{r}_{j} \triangleq [x_{e}^*,y_{e}^*]^{\top}$ satisfying $f(\mathbf{r}_{j}) = 0$ for $j \in \mathbb{Z}_{[1,N_{r}]}$, where $f(\cdot)$ is the function that represents the desired reference.
  
For the investigated switched system \eqref{eq:switched_system}, the switching sequences are required to be admissible, adhering to dwell-time and mode transition restrictions. We begin by defining dwell time of a mode $m \in \mathcal{M}$ as the minimal time between switches \cite{bridgeman2016stability}: 
\[
	\textrm{dwell}_{m}(\sigma) = \min\{ \tau_{s+1} - \tau_{s} \,|\, \sigma(\tau_{s}) = m , s \in \mathbb{N} \}, 
\]
where the switching instant $\tau_{s}\in \mathbb{N}$ is the discrete time when the switching sequence changes mode $\sigma(\tau_{s}) \neq \sigma(\tau_{s}-1)$.

With given reference $\bar{\mathbf{r}}$, the minimum dwell-time for each mode i.e., $d_m$ can be inferred, and $\textrm{dwell}_{m}(\sigma) \geq d_{m}$ holds. In each mode $\sigma(k)$, the remaining dwell-time $\delta(k)$ is computed as
		\begin{equation*}
			\delta(k+1)=
			\begin{cases}
				\max \{ \delta(k)-1, 0 \}, & \text{if }  \sigma(k+1) = \sigma(k), \\
				d_{\sigma(k+1)}, & \text{otherwise.}
			\end{cases}
		\end{equation*}

The condition $\textrm{dwell}_{m}(\sigma) \geq d_{m}$ ensures that the switched system~\eqref{eq:switched_system} in each mode $m \in \mathcal{M}$ can be effectively stabilized using linear controllers of the form $u(k) = K_{m}x(k)$, where $K_m \in \mathbb{R}^{n_{u} \times n_{x}}$ represents the controller gain \cite{liberzon2003switching}. 

\begin{remark}
		In practice, the time of system that dwells in each mode may be unknown, but it is possible to infer a lower bound $d_{m}$ for the dwell time of each mode $m$ based on the information from given reference $\mathbf{r}(k)$. The lower bound always exists considering the special case $d_{m} =1$ when system mode switches. This lower bound signifies that the system is expected to remain in mode $m$ for at least $d_{m}$ time steps.
\end{remark}

\begin{assumption}
    Given a known reference, the minimum dwell time $d_{m}$ of each mode $m\in \mathcal{M}$ can be pre-determined for the design of the controller.
\end{assumption}

    For contouring applications, the admissible mode transition is enforced when a reference is given. This admissible mode transition can be represented by a graph $\mathbb{G} = (\mathcal{M},\mathcal{E})$. The graph nodes represent the modes $\mathcal{M}$ of the switched system \eqref{eq:switched_system} and the edge $(m,n) \in \mathcal{E}$ means the mode transition from $\sigma(\tau_{s}) = m$ to $\sigma(\tau_{s+1}) =  n$ is allowed.

The switching sequence $\sigma$ is drawn from the set:
\begin{equation*}
    \Sigma= \{ \sigma: \mathbb{N} \rightarrow \mathcal{M} \,|\, 
			\left(\sigma(\tau_{s}),\sigma(\tau_{s+1})\right) \in \mathcal{E},\forall s \in \mathbb{N}\}. 
\end{equation*}

\subsection{Bounded contouring error}
		
\begin{figure}[!t]\centering
	\includegraphics[width=\hsize]{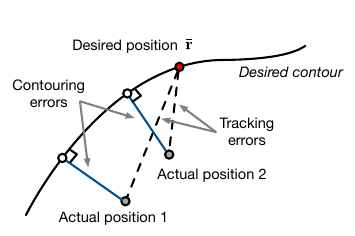}
	\caption{Contouring errors and tracking errors.}\label{fig:err_contour}
\end{figure}

As shown in Fig.~\ref{fig:err_contour}, the contouring error $\epsilon(x_{e},y_{e})$ by its definition is the shortest distance between the actual position of the end effector and the desired contour, while the tracking error is the difference between the actual position and its time-dependent reference.

Recognizing that contouring error significantly impacts product surface quality, our objective is to achieve bounded contouring control in manufacturing processes. It is essential to note that, in certain applications, meeting predefined contouring error tolerances may be more critical than striving for absolute minimization. Hence, our focus is on maintaining contouring error within specified tolerances to ensure high-quality manufacturing outcomes, as denoted by
\begin{equation}
	\epsilon (x_{e},y_{e}) \leq \epsilon_{c}, \label{eq:bounded_err}
\end{equation}
where $\epsilon_{c} \geq 0 $ is a given tolerance of contouring error.

With a given trajectory reference $\bar{\mathbf{r}}$ and admissible switching sequence $\sigma \in \Sigma$, this work is dedicated to solve the following problem:

\begin{problem}\label{prob:prob_thiswork}
For biaxial systems in the form of~\eqref{eq:switched_system} and reference $\bar{\mathbf{r}}$ with a given tolerance $\epsilon_{c}$, design a contouring control law $\mathbf{u}(k) = \kappa(\mathbf{r},\mathbf{x},\sigma,\delta)$ such that the output follows the reference $\bar{\mathbf{r}}$ while the system constraints \eqref{eq:cons_sys} and contouring error bound \eqref{eq:bounded_err} are always satisfied.
\end{problem}

\section{Switched Model Predictive Controller for Bounded Contouring Error}\label{sec:controller}

In this section, we present a control framework based on an MPC controller that effectively ensures bounded contouring error for a switched linear system. It is followed by the analysis of the closed-loop properties of the system.

\subsection{Computation of switch control invariant set}

We start by defining a feasible set of states that satisfies state and contouring error constraints as
\begin{equation}\label{eq:set_S}
	\mathcal{S} = \{\mathbf{x}\in \mathbb{R}^{n_x} \,|\, \mathbf{x} \in \mathcal{X}, \epsilon(x_{e},y_{e}) \leq \epsilon_{c}\}.
\end{equation}
  
In the following, we introduce two important set definitions that will be used to construct the switch CI sets. 

\begin{definition}[Control Invariant Set] 
For each individual mode $m \in \mathcal{M}$, a set $\mathcal{R}_{m} \subseteq \mathcal{X}$ is called control invariant set for the system $\mathbf{x}(k+1) = A_{m}\mathbf{x}(k)+B_{m}\mathbf{u}(k) $ if 
   \begin{equation*}
       \exists \mathbf{u}(k) \in \mathcal{U}, A_{m}\mathbf{x}(k) + B_{m}\mathbf{u}(k)\in \mathcal{R}_{m},\forall \mathbf{x}(k) \in \mathcal{R}_{m}, k \in \mathbb{N}.
   \end{equation*}
\end{definition}

Then a modified backward reachable set is defined as follows:
\begin{definition}[Backward Reachable Set]\label{def:preset}
The set of all states $\mathbf{x} \in \mathcal{S}$ that can be steered into a given set $\mathcal{I} \in \mathbb{R}^{n_{x}}$ in one step under dynamics pair $(A_{m}, B_{m})$ is the backward reachable set (BRS) of $\mathcal{I}$ under mode $m\in \mathcal{M}$ :
\begin{equation*}
\mathsf{B}_{m}(\mathcal{S},\mathcal{I}) = \{ \mathbf{x}\in \mathcal{S} \, | \,\exists  \mathbf{u}\in \mathcal{U}, A_{m}\mathbf{x} + B_{m}\mathbf{u}  \in \mathcal{I}\}.
\end{equation*}
\end{definition}
		
By Definition \ref{def:preset}, the $i$-step BRS of set $\mathcal{I}$ under mode $m$ is defined as 
    \begin{equation*}
	\mathsf{B}_{m}^{i}(\mathcal{S},\mathcal{I})=
		\begin{cases}
			\mathcal{I}, & i=0,\\
			\mathsf{B}_{m}(\mathcal{S},\mathsf{B}^{i-1}_{m}(\mathcal{S},\mathcal{I})), & i\in \mathbb{Z}_{+}.
		\end{cases}
    \end{equation*}

For each mode $m \in \mathcal{M}$, a CI set $\mathcal{R}_{m} \subseteq \mathcal{S}$ can be computed based on the set propagation algorithm in~\cite{yuan2019error}. For a linear time-invariant system, state and contouring error constraints can be satisfied by enforcing the system state within the computed CI set. For cases with mode switches $(m,n) \in \mathcal{E}$, we adapt the idea of switch control invariant sets in \cite{danielson2019necessary} and extend it for satisfying contouring error bound. The definition of switch CI sets is given in the following.
		
\begin{definition}[Switch Control Invariant Sets]

The sets $\mathcal{C}_{m}$, $ \forall m \in \mathcal{M}$, are said to be switch control invariant if each set $\mathcal{C}_{m}$ is control invariant, and for each mode $m \in \mathcal{M}$ with any admissible switch $(m,n) \in \mathcal{E}$ it also holds $\mathcal{C}_{m} \subseteq \mathrm{Pre}_{n}^{d_{n}}(\mathcal{C}_{n})$, where $d_{n}$ is the minimum dwell-time of mode $n$.

\end{definition}

These switch CI sets are essential in subsequent MPC formulation in ensuring the constraints satisfaction for any admissible switching $\sigma \in \Sigma$. The detailed steps of computing these sets are summarized in Algorithm~\ref{alg:CI set computation}, where the computation is initialized with feasible set $\mathcal{S}$.

\begin{algorithm}[H]
	\caption{Switch Control Invariant Sets Computation}\label{alg:CI set computation}
	\begin{algorithmic}[1]
		\STATE Initialization: $i \leftarrow 0$, $\mathcal{I}_{m}^{0} \leftarrow S$, $\forall m \in \mathcal{M}$
		\WHILE{TRUE}
		\STATE $\mathcal{I}_{m}^{i+1} \leftarrow \mathcal{I}_{m}^{i} \cap \mathsf{B}_{m}(\mathcal{I}_{m}^{i}) \underset{(m,n)\in \mathcal{E}}{\cap} \mathsf{B}_{n}^{d_{n}}(\mathcal{S},\mathcal{I}_{n}^{i})$, $\forall m \in \mathcal{M}$ 
		\IF {$\mathcal{I}_{m}^{i+1} = \mathcal{I}_{m}^{i},\, \forall m \in \mathcal{M}$}
        \RETURN $\mathcal{C}_{(m,n)} \leftarrow \mathcal{I}_{m}^{i}$
        \STATE \textbf{break}
		\ELSE
		\STATE $i \leftarrow i+1$
		\ENDIF
		\ENDWHILE
	\end{algorithmic}
	\label{alg1}
\end{algorithm}

Based on known reference $\bar{\mathbf{r}}$, the switch CI sets $\{\mathcal{C}_{(m,n)}\}_{m\in \mathcal{M}}$ can be computed using Algorithm \ref{alg:CI set computation} and stored offline.

\subsection{Model Predictive Controller Formulation}

At each time instant $k$, the value of $\bar{\mathbf{r}}_{j}$ is assigned to $\mathbf{r}(k)$ if $j\leq N_{r}$ to determine a time dependent reference. Let $\sigma(\tau_{s}) = \sigma(k)$ stands for the mode at time instant $k$, and $\sigma(\tau_{s+1})$ be the subsequent mode. The time dependent reference $\mathbf{r}(k)$, current system mode $\sigma(\tau_{s})$, subsequent mode $\sigma(\tau_{s+1})$ and remaining dwell-time $\delta(k) \geq 0$ are assumed to be known at time instant~$k$. 

For the case that the reference may not be admissible, the proposed controller should steer the system to the closest admissible reference by introducing a pair of steady output and input $(\mathbf{y}_{s},\mathbf{u}_{s})$ as:
\begin{subequations}\label{eq:xs_us}
		\begin{align}
			(\mathbf{y}_{s}(k),\mathbf{u}_{s}(k)) & = \underset{\mathbf{y}_{s},\mathbf{u}_{s}}{\arg\min}\left\Vert \mathbf{r}(k)-\mathbf{y}_{s}\right\Vert _{Q_{s}}^{2},\\
			\text{subject to} & \;\; \mathbf{x}_{s}=A_{\sigma(k)}\mathbf{x}_{s}+B_{\sigma(k)}\mathbf{u}_{s},\\
			& \;\; \mathbf{y}_{s} = C_{\sigma(k)} \mathbf{x}_{s}, \\
			& \;\; (\mathbf{x}_{s},\mathbf{u}_{s})\in (\mathcal{X} \times \mathcal{U}),
		\end{align}
	\end{subequations}
where $Q_{s} \succ 0$ is the tuning parameter to penalize the tracking error.

Then, the MPC controller is designed as following to solve the Problem~\ref{prob:prob_thiswork}:
\begin{subequations}\label{eq:MPC_problem}
	\begin{align}
		J^{*}(k)= & \underset{\mathbf{u}}{\min} \sum_{i=0}^{N-1}\left(\left\Vert \mathbf{y}(i|k)-\mathbf{y}_{s}\right\Vert _{Q}^{2}  +\left\Vert \mathbf{u}(i|k) - \mathbf{u}_{s}\right\Vert _{R}^{2}\right)\nonumber\\
		& \qquad +\left\Vert \mathbf{y}(N|k)- \mathbf{y}_{s}\right\Vert _{P_{\sigma(k)}}^{2} ,\\
		\intertext{subject to}\; & \mathbf{x}(i+1|k)=A_{\sigma(k)} \mathbf{x}(i|k)+B_{\sigma(k)} \mathbf{u}(i|k),\label{eq:MPC_system} \\ 
		& \mathbf{y}(i|k)=C_{\sigma(k)} \mathbf{x}(i|k), \label{eq:MPC_output} \\
		& \mathbf{u}(i|k) \in \mathcal{U}, i \in \mathbb{Z}_{[0,N-1]},\label{eq:MPC input constraint}\\
		& \mathbf{x}(i+1|k) \in \mathcal{S}, i \in \mathbb{Z}_{[0,\delta(k))},\label{eq:MPC state constraint}\\
		& \mathbf{x}(i+1|k)  \in \mathcal{C}_{(\sigma(\tau_{s}),\sigma(\tau_{s+1}))}, i \in \mathbb{Z}_{[\delta(k),N-1]}, \label{eq:MPC CI constraint}\\ 
		& \mathbf{x}(0|k) = \mathbf{x}(k) \label{eq:MPC_initial},
	\end{align} 
\end{subequations}
where \eqref{eq:MPC_system} and \eqref{eq:MPC_output} are the system model that are used to predict the system evolution along the MPC prediction horizon; \eqref{eq:MPC input constraint} is the input constraint; \eqref{eq:MPC state constraint} requires the state to stay in CI set to employ state and contouring error constraints when system mode does not change; \eqref{eq:MPC CI constraint} guarantees the feasible set constraints during mode transition; \eqref{eq:MPC_initial} updates the system states with latest measurement. The matrices $Q, R \succ 0 $ are the tuning parameters that are used to penalize the tracking error and control effort, respectively. 

The optimal solution of the online optimization problem~\eqref{eq:MPC_problem} is denoted by $\mathbf{u}^{*}(k) = \{\mathbf{u}^{*}(0|k),\cdots,\mathbf{u}^{*}(N-1|k)\}$ and the first element $\mathbf{u}^{*}(0|k)$ is applied on the plant at each time instant.

With given parameters $Q$ and $R$, the value of $P_{\sigma(k)}$ is chosen based on the following assumption.

\begin{assumption}\label{ass:P_matrix}
	Given $P_{\sigma(0)} \succ 0 \in \mathbb{R}^{n_{x}\times n_{x}}$, there exist matrices $P_{\sigma(k)}$, $k \in \mathbb{N}$ such that
	\begin{multline}
		P_{\sigma(k)} - (A_{\sigma(k)}+B_{\sigma(k)}K_{\sigma(k)})^{\top}P_{\sigma(k+1)}(A_{\sigma(k)}+B_{\sigma(k)}K_{\sigma(k)}) \\
		-Q - K_{\sigma(k)}^{\top}RK_{\sigma(k)} \succeq 0.
	\end{multline}
\end{assumption}

From Assumption \ref{ass:P_matrix}, it can be seen that for a given initial matrix $P_{\sigma(0)}$, the matrices $P_{\sigma(k)}$ can be iteratively computed. The process of running the proposed MPC is summarized in Algorithm~\ref{alg:MPC-running}.
\begin{algorithm}
	\caption{Contouring Error Bounded MPC for Switched Linear System}
	\label{alg:MPC-running} 
	\begin{algorithmic}[1]
		\STATE Based on a given contouring error bound $\epsilon_{c}$ and desired reference $f(\mathbf{r}_j) = 0$ for $j \in \mathbb{Z}_{[1,N_{r}]}$, compute the feasible set $\mathcal{S}$
		\STATE Compute the switch CI sets based on Algorithm~\ref{alg:CI set computation}
            \STATE Initialization: $k\leftarrow 0$, $ j \leftarrow 1 $
		\WHILE {$k \geq 0$ and $j \leq N_r$}
            \STATE $ \mathbf{r}(k) \leftarrow \mathbf{r}_j $
            \STATE Compute admissible pair $(\mathbf{y}_{s},\mathbf{u}_{s})$ based on optimization problem \eqref{eq:xs_us}
            \STATE Measure $\mathbf{x}(k)$ and set $\mathbf{x}(0|k) = \mathbf{x}(k)$
            \STATE Update $\sigma(k)$ based on $\mathbf{x}(k)$
		\STATE Update $P_{\sigma(k)}$ based on Assumption~\ref{ass:P_matrix}\label{step:solve_optimization}
		\STATE Solve the optimization problem \eqref{eq:MPC_problem} to obtain $\mathbf{u}^{*}(0|k)$
		\STATE Apply $\mathbf{u}^{*}(0|k)$ to the plant 
            \IF{$j < N_r$}
                \STATE $ j \leftarrow j+1 $
            \ENDIF
            \STATE $k \leftarrow k+1$
		\ENDWHILE
	\end{algorithmic}
\end{algorithm}

Note that in Step 1 of Algorithm~\ref{alg:MPC-running}, the feasible set may become non-compact in practical applications, necessitating appropriate set approximation. The detailed procedure for computing the approximated feasible set will be discussed in the Section~\ref{sec:synthesis}.

\subsection{Closed-loop properties}

We next discuss the closed-loop property of the switched biaxial system \eqref{eq:switched_system} operated by the proposed MPC controller~\eqref{eq:MPC_problem}. The theoretical result regarding recursive feasibility is given in the following theorem. 

\begin{theorem}[Recursive Feasibility]\label{theorem:RF}
	Consider Assumptions \ref{ass:stab_obs}-\ref{ass:P_matrix} hold. Given a feasible initial state $\mathbf{x}(0)$, the closed-loop system of \eqref{eq:switched_system} with the MPC controller~\eqref{eq:MPC_problem} is recursively feasible for time-varying reference signal $\mathbf{r}(k), \; k \in \mathbb{N}$.
\end{theorem}

\begin{proof}
Since the optimization problem \eqref{eq:MPC_problem} has a solution at time instant $k$, there exists a feasible control input sequence $\{ \mathbf{u}^*(0|k),\cdots,\mathbf{u}^*(N-1|k) \}$ that generates state trajectories $\{\mathbf{x}^*(1|k),\cdots,\mathbf{x}^*(\delta(k)|k) \in \mathcal{S}\}$ and $\{ \mathbf{x}^*(\delta(k)+1|k),\cdots,\mathbf{x}^*(N|k) \} \in \mathcal{C}_{(\sigma(\tau_{s}),\sigma(\tau_{s+1}))}$. This solution is used to construct a feasible solution at $k+1$. Here, two cases depending on whether mode switch occurs are discussed.

(1) Case 1: $\sigma(k+1) = \sigma(k)$. The input sequence can be chosen as $\mathbf{u}(i|k+1) = \mathbf{u}^*(i+1|k) \in \mathcal{U}$, for $i = 0,\cdots,N-2$, which is feasible. With remaining dwell-time as $\delta(k+1) = \delta(k)-1$, the states $\mathbf{x}(i+1|k+1) = \mathbf{x}^*(i|k) \in \mathcal{S}$ are feasible for $i\in \mathbb{Z}_{[0,\delta(k+1))}$. And states $\mathbf{x}(i+1|k+1) = \mathbf{x}^*(i|k) \in \mathcal{C}_{(\sigma(\tau_{s}),\sigma(\tau_{s+1}))}$ are feasible for $i \in \mathbb{Z}_{[\delta(k+1),N-1]}$ considering $\sigma(k+1) = \sigma(k)$. Since the state $\mathbf{x}(N-1|k+1) = \mathbf{x}^*(N|k) \in \mathcal{C}_{(\sigma(\tau_{s}),\sigma(\tau_{s+1}))}$ and based on the property of the switch CI set, there exists a feasible control input $\mathbf{u}(N-1|k+1) \in \mathcal{U}$ such that $\mathbf{x}(N|k+1) = A_{\sigma(k)}\mathbf{x}(N-1|k+1) +B_{\sigma(k)}\mathbf{u}(N-1|k+1)  \in \mathcal{C}_{(\sigma(\tau_{s}),\sigma(\tau_{s+1}))}$. Thus, problem \eqref{eq:MPC_problem} is feasible at $k+1$ for $\sigma(k+1) = \sigma(k)$.

(2) Case 2: $\sigma(k+1) \not= \sigma(k)$. This implies that there is a mode switch at time $k+1$. The remaining dwell-time changes from $\delta(k) = 0$ to $\delta(k+1) = d_{\sigma(k+1)}$. Let $\mathcal{C}_{(\sigma(\tau_{s+1}), \sigma(\tau_{s+2}))}$ be the switch CI set of mode $\sigma(k+1)$, then we have $\mathbf{x}(0|k+1)=\mathbf{x}^*(1|k)\in \mathcal{C}_{(\sigma(\tau_{s}), \sigma(\tau_{s+1}))} \subseteq \mathsf{B}_{\sigma(k+1)}^{d_{\sigma(k+1)}}\left(\mathcal{S},\mathcal{C}_{(\sigma(\tau_{s+1}),\sigma(\tau_{s+2}))}\right)$. Based on the definition of BRS, this means there exists a sequence of feasible control input $\mathbf{u}(i|k+1) \in \mathcal{U}$ that can generate $\mathbf{x}(i+1|k+1) \in \mathcal{S}$ for $i \in \mathbb{Z}_{[0,\delta(k+1)]}$. Since $\mathbf{x}(\delta(k+1)+1|k+1) \in \mathcal{C}_{(\sigma(\tau_{s+1}), \sigma(\tau_{s+2}))}$ and based on the fact that $\mathcal{C}_{(\sigma(\tau_{s+1}), \sigma(\tau_{s+2}))}$ by itself is a switch CI set for mode $\sigma(k+1)$, there exists feasible inputs $\mathbf{u}(i|k+1) \in \mathcal{U}$ that can generate $\mathbf{x}(i+1|k+1) \in \mathcal{C}_{(\sigma(\tau_{s+1}), \sigma(\tau_{s+2}))}$ for $i \in \mathbb{Z}_{[\delta(k+1),N-1]}$. Thus, optimization problem \eqref{eq:MPC_problem} has a feasible solution at time $k+1$, and the optimal control problem is recursively feasible. 

\end{proof}

We next investigate contouring error boundedness and closed-loop stability. The theoretical results are summarized in the following theorem.
 
\begin{theorem}[Closed-loop Stability]\label{theorem:Stability}
	Consider Assumptions \ref{ass:stab_obs}-\ref{ass:P_matrix} hold. The system \eqref{eq:switched_system} with the MPC controller~\eqref{eq:MPC_problem} is
	\begin{enumerate}
		\item[(i)] stable and the contouring error is bounded for every reference $\mathbf{r}(k) = \mathbf{r}_{j} $ with $ 1  \leq j < N_r$;
		\item[(ii)] asymptotically stable to the admissible output $\mathbf{y}_{s}$ corresponding to the last reference point $\mathbf{r}(k) = \mathbf{r}_{N_{r}}$, $\forall k \in \mathbb{N}$.
	\end{enumerate}
\end{theorem}

\begin{proof}
	(i) This proof is implied by recursive feasibility. From Theorem~\ref{theorem:RF}, starting with a feasible initial state $\mathbf{x}(0)$, for any reference $\mathbf{r}(k)$, the closed-loop system~\eqref{eq:switched_system} with the MPC controller~\eqref{eq:MPC_problem} is always feasible, which indicates at every time $k$ all the constraints of~\eqref{eq:MPC_problem} are satisfied. Therefore, we can know the closed-loop state $\mathbf{x}(k) \in \mathcal{S}$, $\forall k \in \mathbb{Z}$. Hence, the closed-loop system is stable and $\epsilon(x_{e}(k),y_{e}(k)) \leq \epsilon_{c}$ by definition of \eqref{eq:set_S}.

	(ii) When current reference becomes $\mathbf{r}(k) =\mathbf{r}_{N_{r}}$, it implies proximity between the actual state and the reference point. Then, mode switch is not required, ensuring that the mode remains unchanged, i.e., $\sigma(k+1) = \sigma(k)$. The MPC problem \eqref{eq:MPC_problem} is equivalent to a tracking MPC problem to a constant point.
	
	As discussed in Theorem \ref{theorem:RF}, the problem \eqref{eq:MPC_problem} is recursively feasible from a time $k$ to time $k+1$. Based on the solutions at time $k$, a sequence of candidate feasible solutions at time $k+1$ can be chosen as the shifted sequence shown in the proof of Theorem \ref{theorem:RF} case 1.
	
	Considering the total MPC cost as the Lyapunov function, we know that $J^*(k)$ is a quadratic function for any time $k \in \mathbb{Z}_{+}$. We also have that $
    J^*(k+1) - J^*(k) \leq J(k+1) - J^*(k)$, where $J^*$ denotes the optimal MPC cost while $J$ denotes the MPC cost with the chosen feasible candidate sequence.
	
	Denote the admissible output and input pair $(\mathbf{y}_{s},\mathbf{u}_{s})$ corresponding to the reference $\mathbf{r}_{N_{r}}$. By Assumption \ref{ass:P_matrix}, we can derive that
	\begin{align*}
		&J(k+1) - J^*(k) \\
        \leq &- \left\Vert C_{\sigma(k)} \mathbf{x}(k)  -\mathbf{y}_{s}\right\Vert _{Q}^{2}  - \left\Vert \mathbf{u}^*(0|k) - \mathbf{u}_{s}\right\Vert _{R}^{2} \\
		\leq &- \left\Vert C_{\sigma(k)} \mathbf{x}(k)  -\mathbf{y}_{s}\right\Vert _{Q}^{2}\\
		\leq & - \alpha_l \left( \left\Vert C_{\sigma(k)} \mathbf{x}(k)  -\mathbf{y}_{s}\right\Vert\right),
	\end{align*}
	where $\alpha_l(\cdot)$ is a class $\mathcal{K}_{\infty}$ function. Thus, according to the above conditions, we can conclude that the closed-loop system is practically asymptotically stable to $\mathbf{y}_{s}$.
\end{proof}

\section{Synthesis for the proposed MPC} \label{sec:synthesis}

In biaxial contouring applications, the desired contour often consists of linear and/or circular paths, sometimes in combination. It is noteworthy, however, that the feasible set $\mathcal{S}$ for these paths may be non-compact. The compactness of constraints in the optimization problem is crucial to ensure the infimum of the objective function is reachable \cite{danielson2019necessary}. To solve this issue, we present the detail of computing the feasible set $\mathcal{S}$ and switch CI sets that can be used in the MPC formulation from a practical implementation viewpoint.

\subsection{Feasible sets for biaxial applications}

We consider two representative contours, as the desired contour can be seen as a combination of linear and circular elements.

\subsubsection{Linear contour}

In Euclidean space, a linear path can be represented by
\begin{equation*}
	f\left(x_{e}^*, y_{e}^*\right) = a x_e^*+by_e^*+c,
\end{equation*}
where $a$, $b$ and $c$ are constant scalars that formulate the characteristics of a linear path. For a given contouring error bound $\epsilon_{c}$, the feasible set of system state with respect to the linear path is represented as
\begin{equation*} 	\label{eq:feasible_set_linear}
	\mathcal{S}_{l} = \left\{ \mathbf{x} \in \mathbb{R}^{n_x}\,\middle| \, |a x_e+by_{e}+c| \leq \epsilon_{c} \sqrt{a^2+b^2}  \right\}.
\end{equation*}

Given the compact nature of this feasible set $\mathcal{S}_{l}$, the switched CI sets can be computed using Algorithm~\ref{alg:CI set computation} by initializing the set $\mathcal{I}_{m}^{0}$ with $S_{l}$.

\subsubsection{Circular contour}
Let the pair of point $(x_{o},y_{o})$ be the center and $R_{c}$ be the radius of desired circular contour. The circular path can be represented by given function 
\begin{equation*}
	f\left(x_{e}^{*}, y_{e}^{*} \right) = \left(x_{e}^{*} - x_{o}\right)^2 + \left(y_e^* - y_{o}\right)^2 - R_{c}^2.
\end{equation*}

With respect to this circular path, the set of admissible system states that satisfy the bounded contouring error is expressed as 
\begin{equation*}\label{eq:feasible_set_circular}
	\mathcal{S}_{c} = \left\{ \mathbf{x} \in \mathbb{R}^{n_{x}} \, \middle| \,  | R_c-\sqrt{(x_e-x_o)^2+(y_e-y_o)^2} | \leq \epsilon_{c}\right\}.
\end{equation*}

\begin{figure}[tbp]
	\centering
	\subfloat[]
	{	\centering
		\label{fig:fea_set_circle}
		\includegraphics[width = 0.9\hsize]{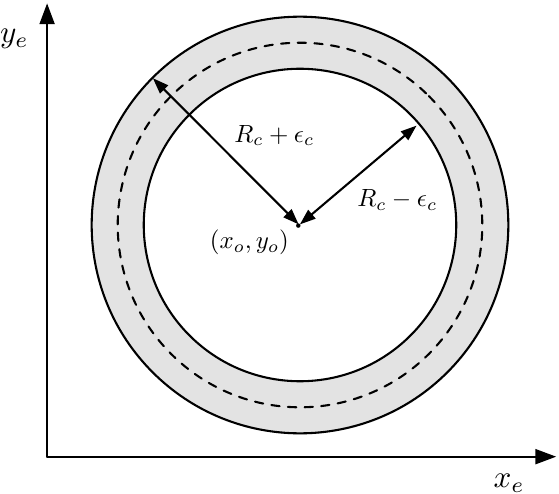}
	}
	\\
	\subfloat[]
	{	\centering
		\label{fig:fea_set_app_circle}
		\includegraphics[width = 0.9\hsize]{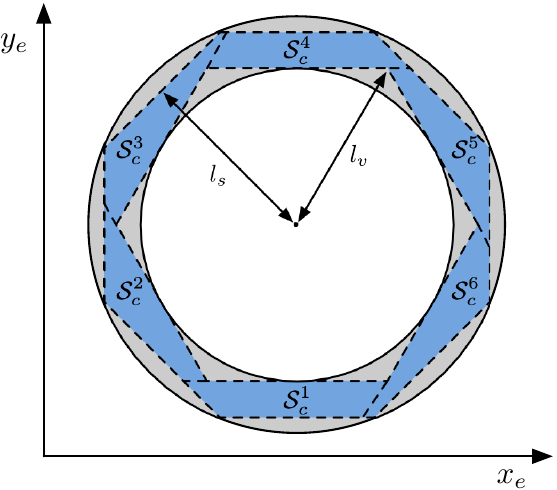}
	}
	\caption{Feasible sets with respect to circular contour: (a) non-compact set $\mathcal{S}_{c}$ (grey shaded area) and desired contour (dot line); (b) approximated sets with partitioned polygons (blue shaded area).\label{fig:fea_set_circle_all}}
\end{figure}

The visual representation of $\mathcal{S}_{c}$ is presented in Fig.~\ref{fig:fea_set_circle}, highlighting its non-compact nature and the necessity for suitable approximation. 

In this paper, we intend to use the polygons for the set approximation. Let $n_{i}$ and $n_{o}$ be the side numbers of inner and outer polygons, respectively. The feasible set $\mathcal{S}_{c}$ can be approximated by the area between the two polygons $\hat{\mathcal{S}}_{c}$, which is partitioned into $n_{i}$ parts of convex sets:
\begin{equation*}
	\hat{\mathcal{S}}_{c} = \underset{p=1}{\overset{n_{i}}{\bigcup}}S_{c}^{p}, \, p \in \mathbb{Z}_{[1,n_{i}]},
\end{equation*}
where each $S_{c}^{p}$ set is compact, and this estimated feasible set $\hat{\mathcal{S}}_{c}$ can be represented by $n_{i}+n_{o}$ numbers of inequalities. Fig.~\ref{fig:fea_set_app_circle} visually illustrates a set approximation example with $n_{i} = 6$ and $n_{o} = 8$.

Let $l_{v}$ and $l_{s}$ denote the distance from the circular center to the vertices of the inner polygon and to the side of the outer polygon, respectively. Once the radius $R_{c}$ and desired contouring error tolerance $\epsilon_{c}$ have been determined, the value of $l_{v}$ and $l_{s}$ are computed as
\begin{align*}
	l_{v} & = \frac{R_c-\epsilon_{c}}{\cos  \left(\pi/n_{i}\right)},\\
	l_{s} & =\left( R_c+\epsilon_{c}\right) \cos\frac{\pi}{n_{o}}.
\end{align*}

To achieve a reasonable approximation of the feasible set, it is essential that the inequalities $l_{v}\leq R_c$ and $l_{v}\leq l_{s}$ are satisfied. Additionally, as the number of inequalities $n_{i}+n_{o}$ directly quantifies the complexity of the approximated feasible set, we aim to solve the following problem to minimize computational complexity:
\noindent
\begin{subequations}
	\begin{align} 
		(\bar{n}_{i}, \bar{n}_{o}) = & \arg \underset{n_{i},n_{o}}{\min}\;  n_{i}+n_{o}, \label{eq:n_in_out_comp}\\ 
		\intertext{subject to}\; & \frac{R_{c}-\epsilon_{c}}{\cos\left(\pi/n_{i}\right)}\leq R_{c} \label{eq:NinCons}, \\
		& \frac{R_{c}-\epsilon_{c}}{\cos\left(\pi/n_{i}\right)}\leq\left(R_{c}+\epsilon_{c}\right)\cos\frac{\pi}{n_{o}}. \label{eq:NoutCons} 
	\end{align}
\end{subequations}

\begin{algorithm}[thbp]
	\caption{Computation of side numbers of inner and outer polygons}\label{alg:Nin_Nout}
	\begin{algorithmic}[1]
		\STATE \emph{Input}: $R_{c}$ and $\epsilon_{c}$
		\STATE \emph{Output}: $\bar{n}_{i}$ and $\bar{n}_{o}$
		\STATE Compute $\bar{n}_{i}$ by solving the optimization problem:
		\begin{align*}
			\bar{n}_{i}=& \arg\min_{n_{i}}  \;n_{i}, \nonumber\\
			\text{subject to} & \;\;\; \frac{R_{c}-\epsilon_{c}}{\cos\left(\pi/n_{i}\right)}\leq R_{c}.
		\end{align*}
		\STATE Compute $\bar{n}_{o}$ by solving the optimization problem:
		\begin{align*}
			\bar{n}_{o}   = & \arg\min_{n_{o}} \;n_{o}, \nonumber\\
			\text{subject to} & \;\;\; \frac{R_{c}-\epsilon_{c}}{\cos\left(\pi/\bar{n}_{i}\right)}\leq\left(R_{c}+\epsilon_{c}\right)\cos\frac{\pi}{n_{o}} .
		\end{align*}
		\IF{$ \{ (x_e^*,y_e^*)|\left(x_e^* - x_{o}\right)^2 + \left(y_e^* - y_{o}\right)^2 = R_{c}^2 \} \subseteq \hat{\mathcal{S}}_{c}$} \label{alg:NinNout_iter}
		\RETURN $\bar{n}_{i}$ and $\bar{n}_{o}$
		\STATE \textbf{break}
		\ELSE
		\STATE $\bar{n}_{o} \gets \bar{n}_{o}+1$
		\STATE \textbf{go to Step} \ref{alg:NinNout_iter}
		\ENDIF
	\end{algorithmic}
\end{algorithm}

Considering that the constraint \eqref{eq:NinCons} is independent of $n_{o}$, a suboptimal algorithm is performed first to determine the minimum $\bar{n}_{i}$ with the given value of $R_c$ and $\epsilon_c$. Then a suitable number $\bar{n}_{o}$ is computed to ensure the existence of approximated feasible sets, and the whole process is summarized in Algorithm~\ref{alg:Nin_Nout}.

After computing the optimal values $\bar{n}_{i}$ and $\bar{n}_{o}$ using Algorithm~\ref{alg:Nin_Nout}, the set $\hat{\mathcal{S}}_{c}$ is determined. This set is then utilized to initialize $\mathcal{I}_{m}^{0}$ in Algorithm~\ref{alg:CI set computation} for the computation of the switch CI set.

\section{Simulation results}\label{sec:application}

To assess the efficacy of the proposed approach, we deploy contouring error-bounded control in a high-fidelity simulation based on a dual-drive industrial laser machine model \cite{yuan2019modelling}. The state vector, denoted as $\mathbf{x}\triangleq [x_{m},y_{N},\theta]^{\top}$, represents the moving distance of end-effector, the moving distance of the mass center of the gantry beam, and the rotation angle of the gantry beam, respectively. The inputs of system consist of the current inputs to the end-effector motor and dual drives. Further details on system identification and model validation can be found in \cite{yuan2019modelling}.

\begin{figure}[tbp]
	\centering
	\subfloat[]
	{	\centering
		\label{fig:ContTraj_XY}
		\includegraphics[width = \hsize]{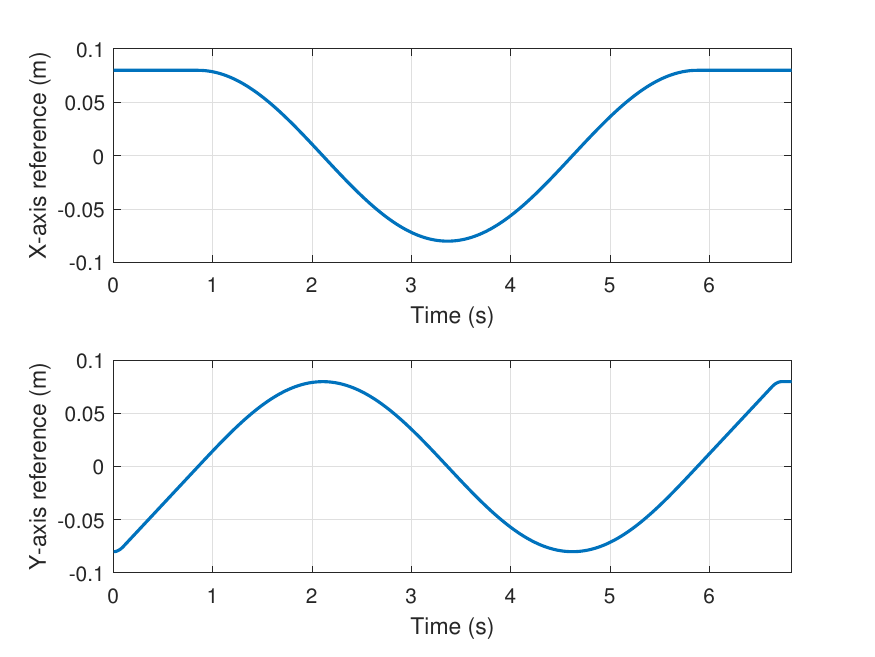}
	}
	\\
	\subfloat[]
	{	\centering
		\label{fig:Cont_XY}
		\includegraphics[width = \hsize]{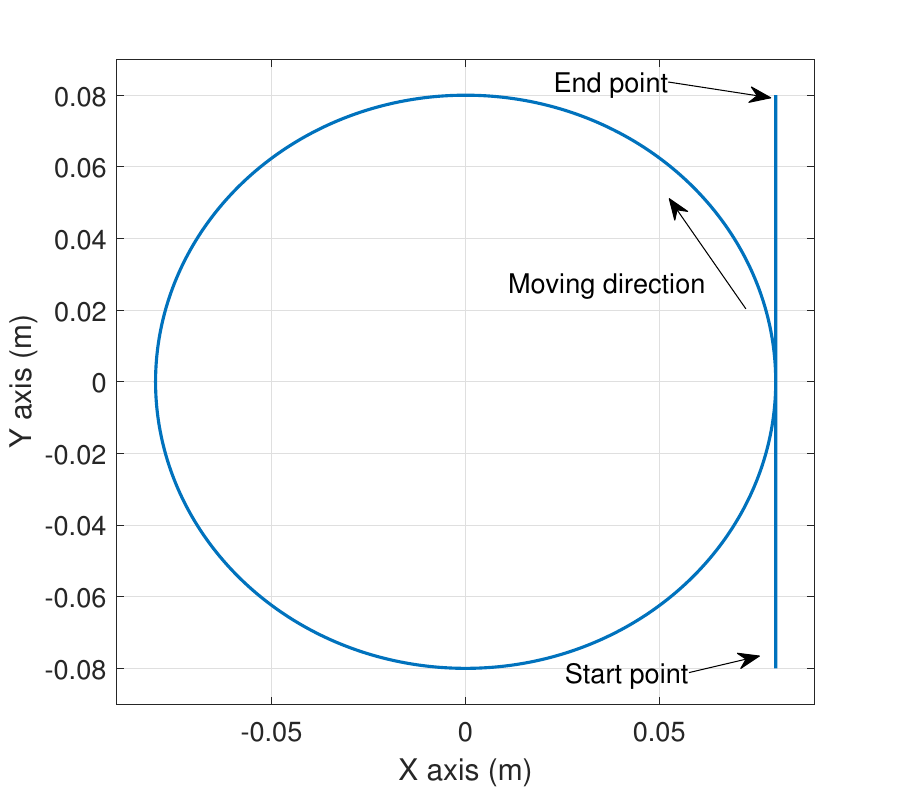}
	}
	\caption{Reference: (a) time-dependent trajectory in X and Y-axis; (b) contour.\label{fig:cont_ref}}
\end{figure}

By using methods such as feedback linearization, the dynamics of this laser machine can be approximated in a form as \eqref{eq:switched_system} where the system matrices are state-dependent on the moving distance of the end-effector along the beam. In the contouring control task, we require the end-effector to follow a contour consisting of a circular path and straight lines with $0.1$ m/s maximum velocity and $1$ m/s$^2$ maximum acceleration. The intended path and trajectory along the X and Y-axes are illustrated in Fig.~\ref{fig:cont_ref}, where the desired circle has a center at $(0,0)$ with a radius of $0.08$ m.

The contouring error bound $\epsilon_{c}$ is chosen as $4$ mm, which corresponds to $5\%$ of the circular radius. The controller update rate and data sampling rate are $500$ Hz. The switched system is divided into five modes based on the linearized operating points at $x_{m} = \{ -0.075, -0.025,\allowbreak 0.025, 0.075 \}$ m to achieve a trade-off between model accuracy and computation effort.

\begin{figure}[tb]
	\centerline{\includegraphics[width = \hsize]{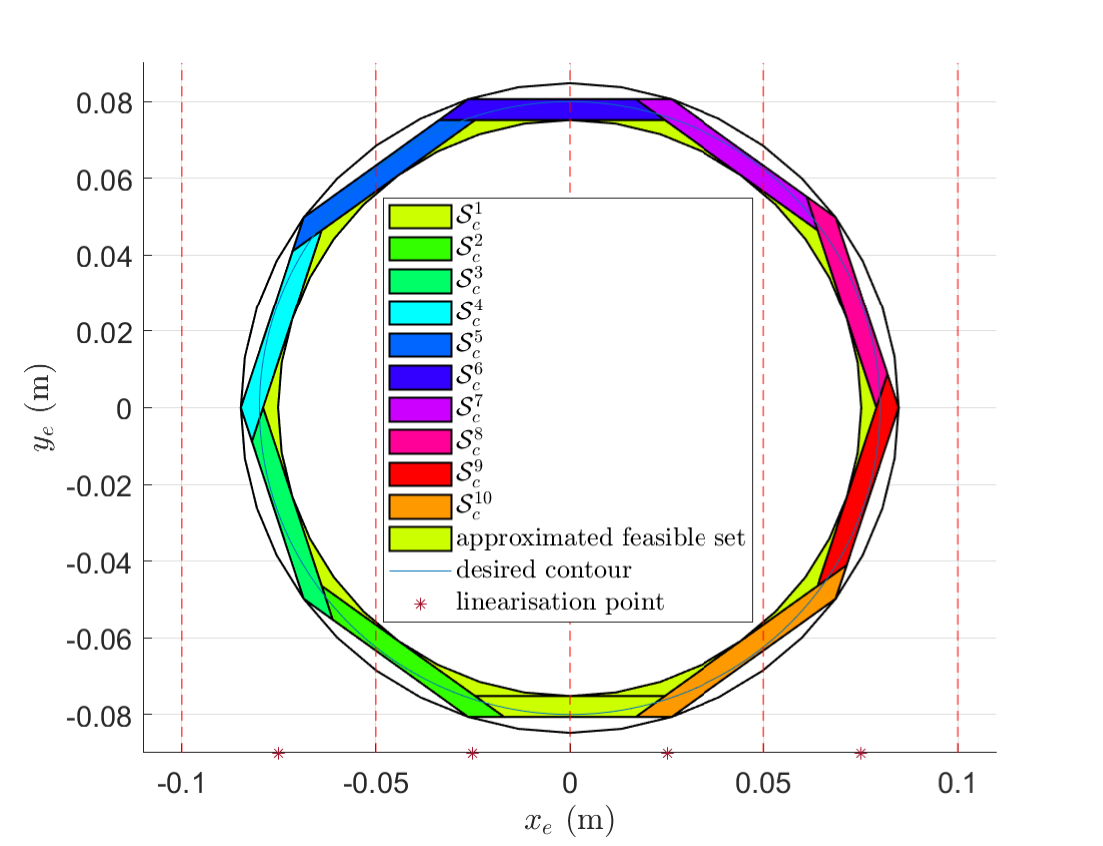}}
	\caption{Feasible sets for bounded contouring error.}{\label{fig:cont_RCI_name}}
\end{figure}

The values of $\bar{n}_{i} = 10$ and $\bar{n}_{o} = 10$ are determined by Algorithm~\ref{alg:Nin_Nout}, and $10$ approximated feasible sets for circular contour are found and shown in Fig.~\ref{fig:cont_RCI_name}.

The prediction horizon in MPC formulation \eqref{eq:MPC_problem} is chosen as $N = 3$, and the tuning weights are chosen as diagonal matrix $Q = \mathrm{diag}(10^{5},10^{5})$ and $R = \mathrm{diag}(10^{-1},10^{-3},10^{-2})$. The value of $P$ is updated based on Assumption~\ref{ass:P_matrix}. 

\begin{figure}[tb]
	\centerline{\includegraphics[width = \hsize]{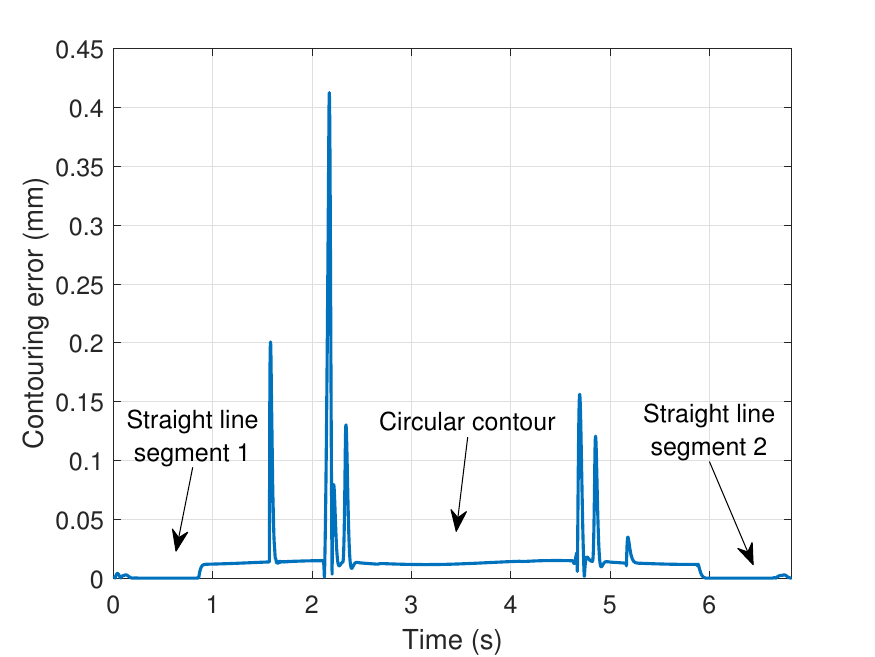}}
	\caption{Contouring error based on proposed control method.}{\label{fig:contour_err}}
\end{figure}

\begin{figure}[tbp]
	\centering
	\subfloat[]
	{	\centering
		\label{fig:track_x_err}
		\includegraphics[width = \hsize]{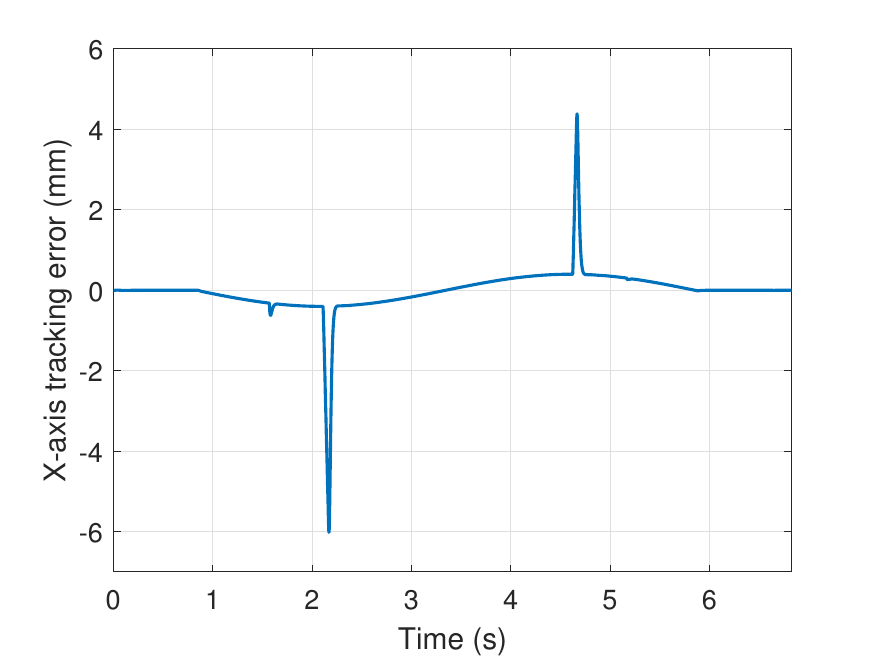}
	}
	\\
	\subfloat[]
	{	\centering
		\label{fig:track_y_err}
		\includegraphics[width = \hsize]{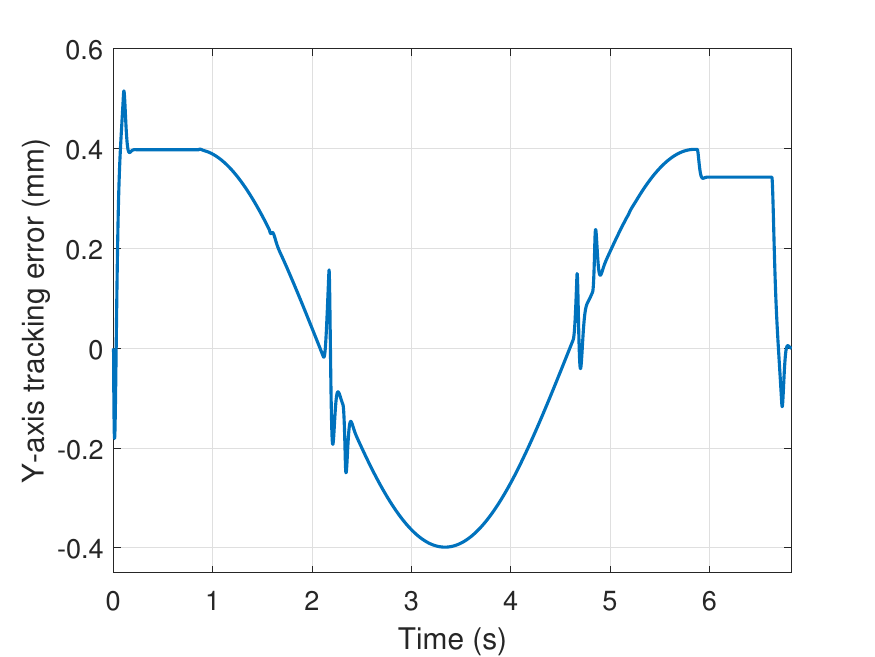}
	}
	\caption{End-effector tracking error: (a) X-axis; (b) Y-axis.\label{fig:track_err}}
\end{figure}  

The entire contouring error throughout the process is computed and depicted in Fig.~\ref{fig:contour_err}. The maximum contouring error is $0.412$ mm, which falls within the acceptable error tolerance. To provide a more comprehensive view of the performance of the proposed controller, the tracking errors along the X and Y-axes are computed and presented in Fig.~\ref{fig:track_err}.

The contouring error graph in Fig.~\ref{fig:contour_err} correlates with the three distinct segments of the path. Within the first $0.85$ s of the straight-line movement, the contouring error remains insignificant due to the limited X-axis displacement. However, during the circular contour segment, a change in the X-axis reference leads to a spike in tracking error, contributing to the contouring error. After $5.87$ s, the straight line becomes the intended contour, with the X-axis tracking error directly influencing the contouring error. Despite a tracking error along the Y-axis, the contouring error stays close to zero due to a minor X-axis tracking error. Notably, even though the maximum tracking error along the X-axis surpasses $6$ mm, the contouring error remains significantly below the desired tolerance. 

In Fig.~\ref{fig:track_err}, spikes in X-axis tracking error occur around time $1.5$ s, $2.1$ s, and $4.6$ s. These spikes, observed by comparing the tracking error with the time-dependent trajectory and linearization points, can be attributed to changes in system modes within the proposed controller. Regarding the Y-axis, the initial sharp tracking error is a consequence of the trade-off between control effort and tracking error when the contouring error is within tolerance. During the circular contour phase, a change in the system mode induces a similar tracking error spike on the Y-axis. After $5.8$ s, the desired contour reverts to a straight line and the Y-axis tracking error begins to decrease as the penalty on tracking error becomes dominant in the cost function. While introducing more system modes could potentially reduce the magnitude of these spikes, it would involve a trade-off between the number of controller switches and the spike magnitude, likely increasing the memory requirements and computational complexity of the proposed algorithm.

\section{Conclusion}\label{sec:conclusion}

In this study, we introduce a contouring error-bounded control algorithm designed for biaxial switched linear systems, and evaluate its effectiveness on an industrial laser machine. The proposed algorithm is able to steer the end-effector to the closest admissible reference. We rigorously discuss the recursive feasibility and stability of the closed-loop system. To ensure the contouring error constraints during mode switch, the states are required to stay within a group of designed switch CI sets. Through theoretical proof and high-fidelity simulations, we affirm that our proposed method can successfully achieve contouring error-bounded control for industrial machines, particularly those with position-dependent structural flexibility. In future work, the proposed method can be extended to the systems with disturbances.

\section*{Acknowledgments}
The authors would like to thank Prof Yutao Chen from Fuzhou University and Prof Iman Shames from Australian National University for their enlightening discussion. We would also like to express our gratitude to the engineers at ANCA Motion for their insightful discussion from a practical perspective.

\bibliographystyle{IEEEtranDOI}
\bibliography{IEEE_ref}

\end{document}